\documentclass{jfm}

\usepackage{graphicx}
\usepackage{natbib}

\ifCUPmtlplainloaded \else
  \checkfont{eurm10}
  \iffontfound
    \IfFileExists{upmath.sty}
      {\typeout{^^JFound AMS Euler Roman fonts on the system,
                   using the 'upmath' package.^^J}%
       \usepackage{upmath}}
      {\typeout{^^JFound AMS Euler Roman fonts on the system, but you
                   dont seem to have the}%
       \typeout{'upmath' package installed. JFM.cls can take advantage
                 of these fonts,^^Jif you use 'upmath' package.^^J}%
      }
  \else
  \fi
\fi

\ifCUPmtlplainloaded \else
  \checkfont{msam10}
  \iffontfound
    \IfFileExists{amssymb.sty}
      {\typeout{^^JFound AMS Symbol fonts on the system, using the
                'amssymb' package.^^J}%
       \usepackage{amssymb}%
         \let\leq=\leqslant
         \let\geq=\geqslant
      }{}
  \fi
\fi

\ifCUPmtlplainloaded \else
  \IfFileExists{amsbsy.sty}
    {\typeout{^^JFound the 'amsbsy' package on the system, using it.^^J}%
     \usepackage{amsbsy}}
    {\providecommand\boldsymbol[1]{\mbox{\boldmath $##1$}}}
\fi

\newcommand\hatR{\skew3\hat{R}}      

\newsavebox{\astrutbox}
\sbox{\astrutbox}{\rule[-5pt]{0pt}{20pt}}

\newcommand\p{\ensuremath{\partial}}

\newcommand\shalf{\ensuremath{{\scriptstyle\frac{1}{2}}}}

\usepackage{graphicx,epsfig,epstopdf,dsfont,color}\usepackage{soul}
\usepackage{amsmath,amssymb,graphicx}
\usepackage{bm,gensymb}
\usepackage[dvipsnames]{xcolor}
\usepackage{float}
\usepackage{xfrac}
\usepackage{url}

\usepackage{array}
\newcolumntype{L}[1]{>{\raggedright\let\newline\\\arraybackslash\hspace{0pt}}m{#1}}
\newcolumntype{C}[1]{>{\centering\let\newline\\\arraybackslash\hspace{0pt}}m{#1}}
\newcolumntype{R}[1]{>{\raggedleft\let\newline\\\arraybackslash\hspace{0pt}}m{#1}}

\definecolor{darkblue}{RGB}{83,0,93}

\def\NA{\nabla}

\def\u{\bm{u}}

\def\ba#1\ea{\begin{align}#1\end{align}}
\def\bsa#1#2\esa{\begin{subequations}\label{#1}
\begin{align}#2\end{align} \end{subequations}}
\def\lp{\left(}
\def\rp{\right)}
\def\lb{\left[}
\def\rb{\right]}

\def\NA{\bm{\nabla}}
\def\DEL{\nabla^2}
\def\f{\frac}

\def\co{\mathcal{O}}
\def\p{\partial}
\def\d{\text{d}}
\def\NA{\nabla}

\title[Internal waves generated by turbulent convection]{The energy flux spectrum of internal waves generated by turbulent convection}

\author[L.-A. Couston, D. Lecoanet, B. Favier, M. Le Bars]{Louis-Alexandre Couston$^1$\thanks{louisalexandre.couston@gmail.com}, Daniel Lecoanet$^{2}$, Benjamin Favier$^1$, Michael Le Bars$^1$}

\affiliation{$^1$ CNRS, Aix Marseille Univ, Centrale Marseille, IRPHE, Marseille, France \\ $^2$ Princeton Center for Theoretical Science, Princeton, NJ 08544, USA}

\begin{document}
\maketitle

\begin{abstract}
We present three-dimensional direct numerical simulations of internal waves excited by turbulent convection in a self-consistent, Boussinesq and Cartesian model of convective--stably-stratified fluids. We demonstrate that in the limit of large Rayleigh number ($Ra\in [4\times 10^7,10^9]$) and large stratification (Brunt-V\"{a}is\"{a}l\"{a} frequencies $f_N \gg f_c$, where $f_c$ is the convective frequency), simulations are in good agreement with a theory that assumes waves are generated by Reynolds stresses due to eddies in the turbulent region \cite[][]{lecoanet2013}. Specifically, we demonstrate that the wave energy flux spectrum scales like $k_{\perp}^4f^{-13/2}$ for weakly-damped waves (with $k_{\perp}$ and $f$ the waves' horizontal wavenumbers and frequencies), and that the total wave energy flux decays with $z$, the distance from the convective region, like $z^{-13/8}$.
\end{abstract}

\vspace{-0.2in}
\section{Introduction}\label{sec:introduction}

Internal gravity waves can be found in planetary atmospheres and oceans, radiative envelopes of stars, and planetary fluid cores. Numerous studies have shown that internal gravity waves can drive strong large-scale flows \cite[][]{grisouard2012,bordes2012,bremer2018,couston2018} and enhance turbulence and mixing \cite[][]{staquet2002,kunze2017,thorpe2018}. Through their effects on mean flows and properties of the fluid in which they propagate, internal waves can have a significant impact on the long-term evolution of global systems, such as Earth's climate \cite[][]{Alexander2010} or the differential rotation of stars \cite[][]{Rogers2012}. 

The energy flux of internal waves is required to accurately predict the influence of waves on global dynamics, but is difficult to estimate, especially when the generating source is turbulent. Nevertheless, turbulent sources account for a significant fraction of the internal wave energy in the atmosphere, stars and planetary interiors, so they can hardly be neglected. Turbulent motions in the bottom Ekman layer \cite[][]{Taylor2007} or the surface mixed layer of the ocean \cite[][]{Munroe2014}, turbulent buoyant plumes in the atmosphere \cite[][]{Ansong2010}, fingering convection \cite[][]{garaud2018} and penetrative convection in stars \cite[][]{Pincon2016} are examples of turbulent sources that can generate internal gravity waves. 

Past studies of turbulence-driven internal waves have focused on extracting scaling laws for the total wave flux with respect to problem parameters. Information on the energy flux spectrum, i.e. the distribution of energy for different wavenumbers $k$ and frequencies $f$ of internal waves, is often beyond reach or difficult to interpret because of experimental or numerical constraints. For instance, the simulations of \cite{Rogers2013} were restricted to only 2D because of this difficulty. However, knowledge of the energy flux spectrum is essential to predicting wave effects, since both wave propagation, damping and interaction with the background state depend strongly on $k$ and $f$.

Several theoretical approaches have been put forward to predict the wave spectrum due to turbulent motions. Under the linearized rapid distortion theory \cite[e.g.][]{carruthers1986}, the wave amplitude is proportional to the convective velocity. We instead assume the wave amplitude is linearly proportional to the Reynolds stress (i.e., the square of the convective velocity), which can accurately predict waves in direct numerical simulations (DNS) \cite[][]{lecoanet2015}. Two theoretical descriptions of the Reynolds stresses in turbulent convection are the ``eddy'' and ``plume'' theories. In the ``eddy'' theory, the convection is decomposed into eddies which follow Kolmogorov statistics and eddy stresses are used as generating sources of internal waves in the stable region \cite[][]{Goldreich1990,lecoanet2013}. In the ``plume'' theory, the convection is seen as an ensemble of uncorrelated plumes impinging at the base of the stable layer \cite[][]{Pincon2016}. The ``plume'' theory predicts the wave energy flux spectrum to be Gaussian in $k$ and $f$; the ``eddy'' theory predicts power laws.

Here we validate for the first time the prediction of the wave flux spectrum based on the ``eddy'' theory in the limit of strong stable stratification \cite[][]{lecoanet2013} using 3D state-of-the-art DNS of a convective--stably-stratified fluid model. The DNS and analyses are challenging as they require a fully turbulent flow interacting with a strongly stratified layer and high spatial/temporal resolution to construct the spectra.

\vspace{-0.1in}
\section{DNS model}\label{sec:model}

\begin{figure}
\centering
\includegraphics[width=\textwidth]{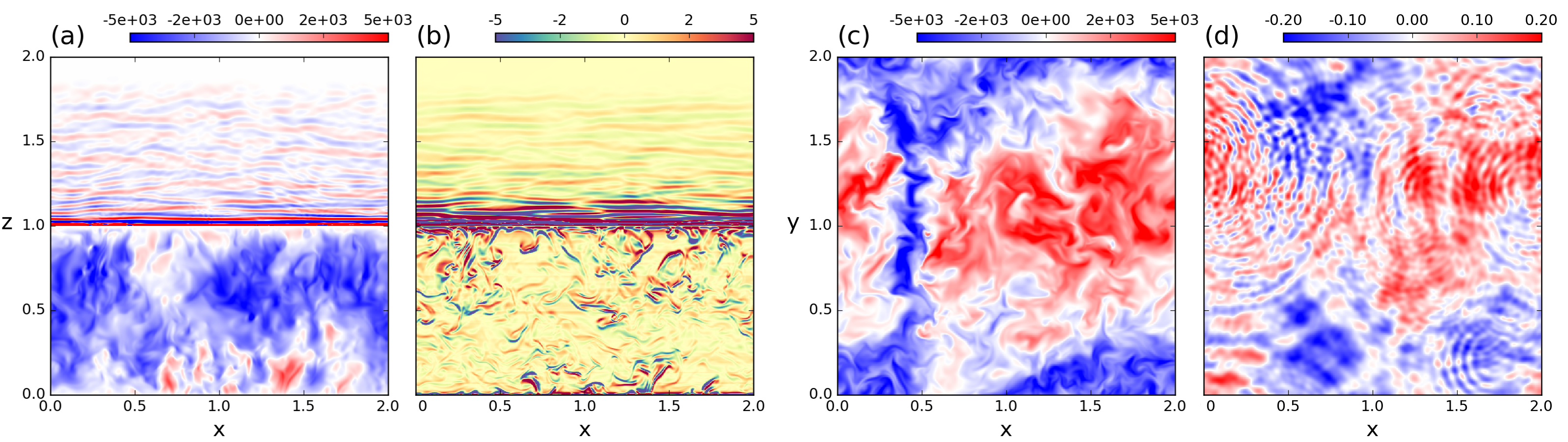}
\vspace{-0.2in}\caption{Snapshots of (a) $w(y=0)$, (b) $T_z-\overline{T}_z$ at $y=0$ (overbar denotes $x$ average), (c) $w(z=0.7)$, (d) $w(z=1.3)$ for case $C_8^{400}$. Variables in the wave region ($z>1$) in (a), (b) have been multiplied by $10^4,10^3$, respectively. A movie is available as Supplementary Material.}\label{fig1}
\end{figure}

The fluid model we use is similar to the one used in  \cite{couston2017}, though here it is 3D (figure \ref{fig1}). We solve the Navier-Stokes equations under the Boussinesq approximation, i.e. the fluid density is $\rho=\rho_0+\delta\rho$ with $\rho_0$ the (constant) reference density and $\delta \rho \ll \rho_0$ are small variations due to temperature fluctuations. We consider constant kinematic viscosity $\nu$, acceleration due to gravity $g$ and thermal diffusivity $\kappa$. The governing equations for the velocity $\u=(u,v,w)$ and temperature $T$ in a Cartesian $(x,y,z)$ frame of reference ($\hat{z}$ is the upward-pointing unit vector of the $z$-axis) read
\bsa{a}
& \p_t \u + (\u\cdot\NA)\u =  - \NA  p + \nu\DEL \u - g  \delta\rho/\rho_0\hat{z}-\u/\tau, \\ 
&\p_t T + (\u\cdot\NA) T = \kappa\DEL T, \\ \label{a4}
&\NA\cdot \u = 0, 
\esa
with $p$ the pressure and $-\u/\tau$ is a damping term used to prevent wave reflection from the top boundary. The density anomaly $\delta\rho$ is related to $T$ through the equation of state
\ba{}\label{b}
\delta\rho/\rho_0=-\alpha(T) (T-T_i)=
-\alpha_s(T-T_i) \text{H}(T-T_i)+ \alpha_s\mathcal{S}(T-T_i)\text{H}(T_i-T),
\ea
with $\text{H}$ the Heaviside function; $\alpha_s>0$ is the thermal expansion coefficient for $T>T_i$, with $T_i$ the inversion temperature at which the density anomaly $\delta\rho/\rho_0=0$ is maximum, and $\mathcal{S}>0$ is the stiffness parameter. We assume periodicity in the horizontal directions and impose free-slip and fixed temperatures conditions on the top ($T=T_t$) and bottom boundaries ($T=T_b$), with $T_b>T_i>T_t$. Then, because equation \eqref{b} is nonmonotonic, the lower part of the fluid is convectively unstable, the upper part is stably stratified and $\mathcal{S}$ indicates how sharp or ``stiff'' the transition is at the interface \cite[][]{couston2017}.

We use the characteristic length scale $H$, the thermal diffusion time $H^2/\kappa$, and the temperature difference $\Delta T=T_b-T_i$ in order to define the non-dimensional variables:
\ba{}\label{c}
(\tilde{x},\tilde{y},\tilde{z})=\f{(x,y,z)}{H},~\tilde{t}=\f{t\kappa}{H^2},~\tilde{\u} = \f{\u H}{\kappa},~\tilde{T}=\f{T-T_i}{\Delta T},~\tilde{p}=\f{p H^2}{\kappa^2},~\tilde{\rho}=\f{\delta\rho}{\rho_0 \alpha_s\Delta T}.
\ea
Substituting \eqref{c} in \eqref{a}-\eqref{b} and dropping tildes we obtain the dimensionless equations
\bsa{d}\label{da}
& \p_t \u + (\u\cdot\NA)\u =  - \NA  p + Pr\DEL \u - Pr Ra  \rho\hat{z} - \u/\tau, \\ 
&\p_t T + (\u\cdot\NA) T = \DEL T, \\ \label{aa4}
&\NA\cdot \u = 0, \\
& \rho= -T\text{H}(T)+\mathcal{S}T\text{H}(-T),
\esa
with $Pr=\nu/\kappa$ the Prandtl number and $Ra=\alpha_sg\Delta T H^3/(\nu\kappa)$ the Rayleigh number. We note $L_x,L_y,L_z$ the dimensionless lengths of the domain in $x,y,z$ directions and $T_{top}<0$ the dimensionless temperature at $z=L_z$ ($T=1$ at $z=0$ and the inversion temperature is $T=0$ in dimensionless space). The dimensionless $z$-dependent damping coefficient used to prevent wave reflection from the top boundary is $\tau^{-1}(z)=0.5 f_{N^*}\{\tanh[ (z-L_z+0.15)/0.05 ]+1\}$ with $f_{N^*}$ the target buoyancy frequency (cf. next section).

We solve \eqref{d} using the open-source pseudo-spectral code DEDALUS \cite[][]{Burns2017}. We are interested in DNS results at statistical steady-state and in the limit of fully turbulent convection (large $Ra$) and strong stable stratification (large $\mathcal{S}$). For ease of presentation, we want the convection to extend from $z=0$ to $z=h \approx 1$ at thermal equilibrium, i.e. such that $H$ is approximately the size of the convection zone in dimensional variables and $Ra$ is the effective Rayleigh number of the convection. $h$ is an output of our simulations, but it can be controlled by iteratively adjusting $T_{top}$ (all other parameters fixed). At leading order, $h\approx 1$ when $T_{top}$ is chosen such that the measured heat flux $Q=\overline{wT-T_z}$ in the convection (overbar denotes the horizontal average) equals the expected diffusive heat flux of the stable region $\sim -T_{top}/(L_z-h)$ with $h= 1$. Because $Q$ in the convection zone changes with $T_{top}$, finding $T_{top}$ is carried out iteratively. We fix all problem parameters except $T_{top}$, make an initial guess for $T_{top}$, and run a simulation for a few convective turnover times, initialized with a temperature of $T=(1-z)\text{H}(1-z)+T_{top}(z-1)\text{H}(z-1)$ plus low-amplitude noise. We then update our guess for $T_{top}$ using $T_{top}=-Q(L_z-h)$ with $h=1$, where $Q$ is the average heat flux at $z=0$, and we reiterate until $T_{top}$ doesn't change by more than about 10\%. The final $T_{top}$ gives an estimate for the buoyancy frequency $f_{N^*}=\sqrt{-PrRa\mathcal{S}T_{top}/(L_z-1)}/(2\pi)$ used in the expression for the damping term $\tau$ in \eqref{da}. The actual buoyancy frequency $f_N=N/(2\pi)$ is computed using $N=\sqrt{PrRa\mathcal{S}Q}$ with $Q$ averaged from $z=0$ to $z=1$ and over the simulation time $\Delta t$. In all the simulations, we set $L_x=L_y=L_z=L$ with $L=2$ or $L=3$. Therefore, the ratio between the horizontal and vertical extent of the convective zone is larger than 2, which ensures that confinement effects associated with the horizontal periodic boundary conditions are not dominating the dynamics. We set $Pr=1$ for simplicity. Input and output parameters of the eight simulations included in this paper are reported in table \ref{table1}.

The variables of interest include the kinetic energy spectrum and the energy flux spectrum. The kinetic energy per unit surface area and unit (thermal) time reads 
\ba{}\label{toto}
K(z) & = \int \f{|\u|^2}{2} \f{\d x \d y \d t}{L^2\Delta t} \approx \sum_{f_i,k_{\perp j}} \f{|\hat{\u}|^2}{2 \delta f \delta k_{\perp}} \delta f \delta k_{\perp}  \equiv \sum_{f_i,k_{\perp j}} d K \delta f \delta k_{\perp} \approx \int \f{\p^2 K}{\p f \p k_{\perp}} \d f\d k_{\perp},
\ea
with hat denoting the Fourier transform in $(x,y,t)$, $\Delta t$ the period over which results are integrated in time, $k_{\perp}$ and $f$ are the horizontal wavenumber and frequency, $\delta k_{\perp}$ and $\delta f$ are the unit increments in spectral space, and $dK$ is the discrete version of the kinetic energy spectrum $\f{\p^2 K}{\p f \p k_{\perp}}$. Similarly, we define the $z$-dependent vertical energy flux as
\ba{}\label{toto2}
F(z) & = \int wp\f{\d x \d y \d t}{L^2\Delta t}  \approx \sum_{f_i,k_{\perp j}} \f{\mathcal{R}(\hat{w} \hat{p}^*)}{ \delta f \delta k_{\perp}} \delta f \delta k_{\perp} \equiv \sum_{f_i,k_{\perp j}} d F \delta f \delta k_{\perp} \approx \int \f{\p^2F}{\p f \p k_{\perp}} \d f\d k_{\perp},
\ea
where $*$ denotes the complex conjugate, $dF$ is the discrete energy flux spectrum and $\mathcal{R}$ denotes the real part. Note that throughout this paper, we assume isotropic motions in $x,y$ directions and express variables in spectral space in terms of the average horizontal wavenumber $k_{\perp}=(k_x^2+k_y^2)^{1/2}$ (with $k_x,k_y$ the wavenumbers in $x,y$ directions). Thus, $|\hat{\u}|^2 \equiv |\hat{\u}(k_{\perp})|^2$ is the azimuthal average of $|\hat{\u}(k_x,k_y)|^2$ and $\mathcal{R}(\hat{w}\hat{p}^*)$ is the azimuthal average of $\mathcal{R}[\hat{w}(k_x,k_y)\hat{p}^*(k_x,k_y)]$ based on the modulus of vector $(k_x,k_y)$, i.e. averaging over all orientations. We apply a Hann windowing function to all variables before taking the time Fourier transform and we discard the modes $f=0$ and $k_{\perp}=0$ from the spectra.

\begin{table}
\centering
	\begin{tabular}{L{0.9cm}R{0.7cm}R{0.8cm}R{0.8cm}C{0.6cm}R{2.7cm}R{0.9cm}R{0.9cm}R{0.9cm}R{0.9cm}R{0.9cm}}
Name & $\f{Ra}{10^8}$ & $\mathcal{S}$ & $T_{top}$ & $L$ & $n_xn_y\times(n_{zl}+n_{zu})$ & $\f{\delta t}{10^{-7}}$ &  $\f{f_c}{10^3}$ &$f_c\Delta t$ & $\f{f_c}{u_{rms}}$ &  $\f{f_N}{f_c}$    \\ \hline	 
$C_7^{10}$ & $0.4$ & 10& -86 & 3 & $256^2\times(384+256)$ & 0.5 & 1.1 & 9.8 & 1.6   & 19 \\	 
$C_7^{100}$ & $0.4$ & 100& -35 & 2 & $256^2\times(384+128)$ & 2 & 1.2 &   10  & 1.5 & 50  \\	 
$C_7^{400}$ & $0.4$ & 400& -32 & 2 & $256^2\times(384+192)$ & 2 & 1.3 &  10.7  & 1.9 & 86 \\	 
$C_8^{10}$ & $2$ & 10& -140 & 3 & $256^2\times(512+256)$ & 0.2  & 2.6 &   7.1  & 1.6 & 23 \\	 
$C_8^{50}$ & $2$ & 50& -62 & 2 & $256^2\times(512+256)$ & 0.6 & 2.5 &   9.1  & 1.5 & 49\\	 
$C_8^{100}$ & $2$ & 100& -60 & 2 & $256^2\times(512+256)$ & 0.6 & 2.4 &  12.5  &  1.4 & 71 \\	 
$C_8^{400}$ & $2$ & 400& -52 & 2 & $512^2\times(768+384)$ & 0.6 & 3.0 &  3.8  & 1.9 & 109\\	 
$C_9^{100}$ & $10$ & 100 & -100 & 2 & $512^2\times(768+256)$ & 0.1 & 5.4 &  5.4 & 1.7  & 90\\ \hline
	\end{tabular}
	\caption{Simulation parameters: number of Fourier modes $n_x,n_y$ in $x,y$ directions, Chebyshev modes $n_{zl}$/$n_{zu}$ in lower/upper regions of the compound $z$  basis (stitched at $z=1.2$), typical time steps $\delta t$, best-fit convective frequency $f_c$ for $\mathcal{A}=30$ (cf. \S\ref{sec:discussion3}),  time $f_c\Delta t$ of data collection, convective velocity $u_{rms}$ and Brunt-V\"{a}is\"{a}l\"{a} frequency $f_N$. $Pr=1$ for simplicity.}	
	\label{table1}
\end{table}
%
%
\vspace{-0.1in}
\section{Theoretical prediction}\label{sec:theory}

We now describe a heuristic model which predicts the wave energy flux spectrum at the interface, $F^{i}(f,k_{\perp})$, between the convective and stably-stratified layers.  This is based on the calculation in \cite{lecoanet2013} (specifically, the ``discontinuous'' $N^2$ case), which itself is based on the earlier calculation in \cite{Goldreich1990}.  We first assume that turbulence driven by the convection can be decomposed into ``eddies,'' each of which has a characteristic length $h$ and velocity $u_h$.  The largest such ``eddy'' is the large-scale circulation of the convection, which has length $\ell_c=2\pi/k_c$ and velocity $u_c$.  We imagine this large scale circulation represents the injection scale of a Kolmogorov turbulent cascade and to recover the $k^{-5/3}$ energy spectrum ($k$ is the three-dimensional wavenumber), we require that the typical velocity of eddies with size $h$ scales like $u_h/u_c=(h/\ell_c)^{1/3}$. We assume each eddy is coherent for an eddy turnover time $\tau_h=h/u_h$, which implies that the power spectrum in the convection zone is peaked along the curve $\tau_h f_c=(h/\ell_c)^{2/3}$. This ignores the fact that small eddies can be advected over their size $h$ on timescales shorter than $\tau_h$ by large-scale motions. This set of assumptions is very simplistic, but allows a theoretical derivation of the wave energy flux spectrum.

In this hypothesis, the waves are generated by Reynolds stresses within the convection zone.  We calculate the waves generated by a single eddy with size $h$ and corresponding turnover time $\tau_h$, and then sum up the wave energy flux over all eddies.  An eddy with size $h$ and turnover time $\tau_h$ can only efficiently excite a wave with frequency $f$ and wavenumber $k_{\perp}$ if both
\ba{}\label{eqn:wave-eddy-link}
f\leq \tau_h^{-1}, \quad \quad \ell_c (k_{\perp}/2\pi) \leq \ell_c/h = (\tau_h f_c)^{-3/2}.
\ea{}
This is because we assume the spectral representation of an eddy's velocity field is peaked at wavenumbers around $2\pi/h$ and decreases exponentially at higher wavenumbers due to analyticity (and similarly for the temporal spectrum with timescale $\tau_h$). Waves with $f>\tau_h^{-1}$ or $k_{\perp}>2\pi/h$ then have exponentially small coupling to eddies of size $h$.

To connect the properties of the convection to the waves, one can write an equation for the wave velocity as a function of turbulent stresses as \cite[][]{lecoanet2013},
\begin{equation}\label{eqn:green}
\nabla^2\partial_t^2w + N^2(z)(\partial_x^2+\partial_y^2)w = S,
\end{equation}
where $S$ is related to the Reynolds stresses associated with convective eddies. Note that we have neglected the effects of diffusion in the excitation process. We also neglect thermal stresses within the convection zone \citep[as in][]{lecoanet2015}. \cite{lecoanet2013} derives the internal waves' Green's function from this equation assuming $N^2(z)$ is constant in the convective and stably-stratified region, and discontinuous at the interface. One can then estimate the convolution of this Green's function with the Reynolds stress of an eddy to find that the wave energy flux from eddies of size $h$ is 
\ba{}\label{eqn:single-eddy-flux}
\frac{\d F^{i}}{\d\log k_{\perp} \d\log f \d\log h}  = \mathcal{A} u_h^3 \frac{f_c}{f_N} \f{f}{f_c}  \lp\f{k_{\perp}h}{2\pi}\rp^4 = \mathcal{A} u_c^3 \frac{f_c}{f_N} \f{f}{f_c} \f{h}{\ell_c} \lp\f{k_{\perp}h}{2\pi}\rp^4,
\ea{}
with $\mathcal{A}$ a multiplicative constant, assumed universal, i.e. independent of all problem, wave or eddy parameters. To calculate the wave energy flux for a given frequency and wavenumber, we must sum up excitation from all eddies which satisfy equation~\eqref{eqn:wave-eddy-link}.  Because of the strong dependence on the eddy size in equation~\eqref{eqn:single-eddy-flux}, the wave energy flux is dominated by the largest possible eddies which can excite a wave. The wave energy flux depends on whether the wave frequency is higher or lower than the frequency of the largest generating eddy $f^*(k_{\perp})=f_c[\ell_ck_{\perp}/(2\pi)]^{2/3}$. Waves with $(k_{\perp},f)$ satisfying $f<f^*(k_{\perp})$ are not observed in the bulk of the stable region in DNS because they are strongly dissipated (see discussion after \eqref{h}), so we neglect their contribution to the energy flux in the theory. The expression for the flux of the waves we keep, i.e. with frequency $f>f_c[\ell_ck_{\perp}/(2\pi)]^{2/3}$, is 
\ba{}\label{eqn:high frequency}
\frac{\d F^{i}}{\d\log k_{\perp} \d\log f} = \mathcal{A} u_c^3 \frac{f_c}{f_N} \left(\frac{f}{f_c}\right)^{-13/2} \lp\f{k_{\perp}\ell_c}{2\pi}\rp^{4}.
\ea{}

We recall that equation \eqref{eqn:high frequency} is a prediction for the energy flux spectrum of \textit{inviscid propagating wave modes} at the convective--stably-stratified interface. In the DNS, however, the energy flux at the interface has contributions from both \textit{inviscid and viscid wave modes} and overshooting \textit{plumes}, such that it is not expected to match the prediction \eqref{eqn:high frequency}. In order to avoid this discrepancy, we compare the theory with DNS away from the interface, where overshooting plumes and strongly-viscous modes can be neglected. 

In order to extend the expression for the flux \eqref{eqn:high frequency} at the interface to an expression valid in all of the stable layer we derive a wave decay rate from the linear equation 
\ba{}\label{h}
(\p_t-\NA^2)^2\NA^2 w+N^2\lp \p_x^2 + \p_y^2 \rp w =0,
\ea
valid for $Pr=1$. Note this differs from the {\it inviscid} equation \eqref{eqn:green} used to derive the wave flux {\it at the interface}. Substituting a plane wave solution $w\sim e^{i(k_xx+k_yy-\omega t)} e^{(ik_z - \gamma)z}$ with $\omega>0$ in \eqref{h} yields a cubic complex equation for $(ik_{z}-\gamma)^2$ whose six roots can be written in closed form. We assume that motions in the stable layer are dominated by waves propagating upward, i.e. with group velocity $C_{gz}=\p\omega/\p k_{z}>0$, and whose amplitude decrease with $z$. One of the three roots $(k_{zj},\gamma_j)$ with $C_{gzj}>0$ has a much smaller decay rate $\gamma_j$ than the other two, such that it is expected to be the dominant contribution to upward-propagating and decaying waves in DNS. Unsurprisingly, this solution is the ``full-dissipation" equivalent of the simpler WKB solution  
\ba{}\label{i}
k_z = \mp k_{\perp} N/\omega, \quad\quad \gamma = \pm k_{\perp}^3 N^3/\omega^4,
\ea
with $k_z<0$ for upward-propagating waves, derived in the weak dissipation limit and valid when $\omega \gg k^2 \geq O(1)$ and $\omega^2 \ll N^2$. We neglect the other two more-rapidly decaying upward-propagating wave solutions of \eqref{h}. Note that whether they capture some of the turbulent energy is beyond the scope of this work, and neglecting them is consistent with prediction \eqref{eqn:high frequency} since the theory only considers the WKB mode \eqref{i} in its inviscid form. Given that a broad range of frequencies and wavenumbers is excited in the DNS, it may be expected that \eqref{i} does not accurately predict the decay rate of the dominant waves. However, we find that the WKB solution and the ``full-dissipation" solution for $\gamma$ yield similar results for the theoretical spectra presented here, so for simplicity we will use \eqref{i} in the remainder of this paper (note that this suggests that the discrepancy between the ``full" and the WKB decay rate is significant only when both decay rates are so large that it affects waves whose amplitudes are so small that they can be neglected). Note for future discussions that the decay rate is strongest for low-frequency high-wavenumber waves, as clearly shown by \eqref{i}.
 
Finally, the $z$-dependent theoretical wave flux spectrum becomes 
\ba{}\label{eq:dFth}
\f{\d F^{th}}{\d \log k_{\perp}\d \log f} = \f{\d F^{i}}{\d \log k_{\perp}\d \log f} e^{-2\gamma(z-z_i)} = \mathcal{A} u_c^3 \frac{2\pi f_c}{N} \left(\frac{f}{f_c}\right)^{-\f{13}{2}} \lp\f{k_{\perp}\ell_c}{2\pi}\rp^{4} e^{-2\gamma(z-z_i)},
\ea
with $\gamma$ given by \eqref{i} and $z_i$ the interface height. A useful approximate expression can be obtained for the total flux as a function of $z$ by integrating \eqref{eq:dFth} in the limit $f_N \gg f_c \gg (z-z_i)k_c^3/\pi$ and sufficiently far from the interface, i.e. $z-z_i \gg [\gamma(k_c,f_c)]^{-1}$. The derivation is tedious but straightforward, yielding (cf. Appendix \ref{appA}) 
\ba{} \label{eq:Fasym}
F^{th} = \int_{f_c}^{f_N} \int_{k_c}^{k_c\lp\f{f}{f_c}\rp^{\f{3}{2}}} \f{\d  F^{th}}{\d  f \d  k_{\perp} }  \d k_{\perp}  \d f \approx \mathcal{A} u_c^3 \f{f_c^{\f{15}{2}}}{f_N^{\f{47}{8}}}  \f{5\Gamma\lp\f{5}{8}\rp \ell_c^{\f{39}{8}}}{28(8\pi^2)^{\f{13}{8}}} \lp z-z_i\rp^{{-\f{13}{8}}} \equiv \tilde{\mathcal{F}}  \lp z-z_i\rp^{{-\f{13}{8}}}.
\ea
The domain of validity of the asymptotic solution is expected to encompass all of the wave region in DNS. In the next section we show that the wave flux spectrum in DNS follows the trends predicted by the theory, i.e. equations \eqref{eqn:high frequency} and \eqref{eq:Fasym}.

\vspace{-0.15in}
\section{Results and discussion}\label{sec:discussion}

\subsection{Statistics of turbulent convection}\label{sec:discussion1}

\begin{figure}
\centering
\includegraphics[width=1\textwidth]{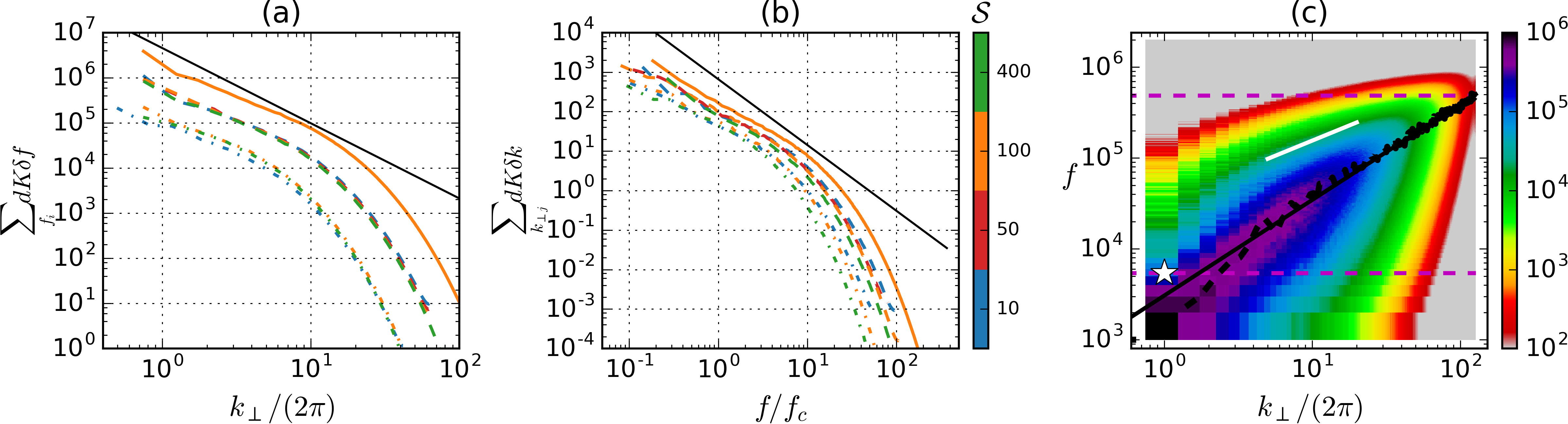}
\vspace{-0.1in}\caption{Integrated kinetic energy at $z=0.7$ as a function of (a) $k_{\perp}$ and (b) $f$. Solid, dash-dash, and dash-dot lines correspond to $Ra=4\times 10^7,2\times 10^8,10^9$ and different colors correspond to different stiffnesses. Black solid lines are power laws with slope -5/3. (c) Kinetic energy spectrum $k_{\perp}fdK$ at $z=0.7$ for $C_9^{100}$. The white star shows the wavenumber-frequency pair $(\ell_c^{-1},f_c)$ of the large-scale circulation that allows a good agreement of the DNS with the theory. The pair $(\ell_c^{-1},f_c)$ is not unreasonably far from the line $f(k_{\perp})$ of maximum energy, which is shown by the black dash-dash line (with solid line showing the linear fit). Dashed purple lines show $f_c,f_N$. The solid white line shows slope $+2/3$.}
\label{fig2}
\end{figure}

The theory described in section \ref{sec:theory} assumes that the spatial correlations of the turbulent convection follow the standard Kolmogorov prediction for homogeneous and isotropic turbulence. Figure \ref{fig2}(a) shows that the frequency-integrated kinetic energy spectrum $\sum_{f_i} dK\delta f$ indeed has a slope relatively close to $k_{\perp}^{-5/3}$ for all simulations, thus validating the use of Kolmogorov hypothesis. The theory also requires prescription of the temporal correlations of the turbulence fluctuations. Here, the theory uses the eddy turnover time as the relevant time scale, which is known as the straining hypothesis. However, one could argue that the temporal correlations of small-scale turbulent features seen at a fixed point in space are affected by the advection by large-scale motions, and that the sweeping hypothesis should be considered instead \cite[][]{tennekes1975,chen1989}. The wavenumber integrated kinetic energy spectrum in \ref{fig2}(b) shows a frequency scaling somewhat weaker than $f^{−5/3}$, which is the typical scaling for turbulent flows in Eulerian frame with sweeping \cite[][]{chevillard2005,canet2017}, in agreement with convection experiments \cite[][]{sano1989}. The straining mechanism would instead lead to a steeper $f^{-2}$ scaling, which is recovered when computing temporal correlations along Lagrangian trajectories in homogeneous turbulence \cite[][]{chevillard2005} and in Rayleigh-B\'enard convection \cite[][]{liot2016}. Figure \ref{fig2}(c) shows $k_{\perp}fdK$ for simulation $C_9^{100}$ in $(k_{\perp},f)$ space, thus providing a more detailed picture of the turbulence statistics of the most turbulent simulation (similar results are obtained for other simulations). Again, it can be seen that the line of maximum $k_{\perp}fdK$ is reasonably well approximated by a power fit of the form $f\sim k_{\perp}^a$ (solid black line), with $a$ close to 1 for all simulations, suggesting strong sweeping effects. Isocontours at higher frequencies, however, have a milder slope $f\sim k_{\perp}^b$ with $b<1$, possibly suggesting that the eddy dominant frequency does scale like $\sim k_{\perp}^{2/3}$ (solid white line) and that sweeping does not affect all eddies. We will show that there is a good agreement between DNS results and the theory based on the straining hypothesis despite the presence of a strong large-scale flow. It should be noted that the theory in \S\ref{sec:theory} changes with the choice of temporal correlations, as is already the case for other wave generation mechanisms by turbulence \cite[for sound waves, see e.g.][]{zhou1996,favier2010}.

\subsection{Integrated wave energy flux}\label{sec:discussion3}

In this section we compare the total energy flux $F(z)$ obtained in DNS with the theoretical prediction $F^{th}(z)$ (cf. \eqref{eq:Fasym}), and we infer the free parameters of the theory $\mathcal{A}$ and $f_c$ from a best fit. The theoretical total flux $F^{th}$ has a steep scaling with the buoyancy frequency $f_N$, the interface height $z_i$ and the turbulence parameters $\ell_c$, $f_c$ and $u_c$. While $f_N$ and $z_i$ are relatively well-constrained, the turbulence parameters are challenging to estimate from DNS of turbulent convection, making $F^{th}$ potentially highly sensitive to the definitions used. A slight variation of $f_c$ by a factor of 2, due to a change of definition, can lead to a change of $F^{th}$ by a factor of $\approx 200$ at all heights. Here, instead of estimating the values of the turbulence parameters using ad-hoc definitions, we use a best-fit approach. For simplicity, we assume that $\ell_c=1$ in all cases and thus take $u_c=f_c$. This leaves $f_c$ as the only free parameter in the theory, besides $\mathcal{A}$, which is assumed to be the same for all simulations. A posteriori, we check that the best-fit $f_c$ is not too different from the rms velocity $u_{rms}=\sqrt{2\mathcal{K}}$ at $z=0.7$ in the convection zone (cf. equation \eqref{toto} and note that $u_{rms}$ is relatively depth-invariant in the convective bulk).

\begin{figure}
\centering
\includegraphics[width=0.95\textwidth]{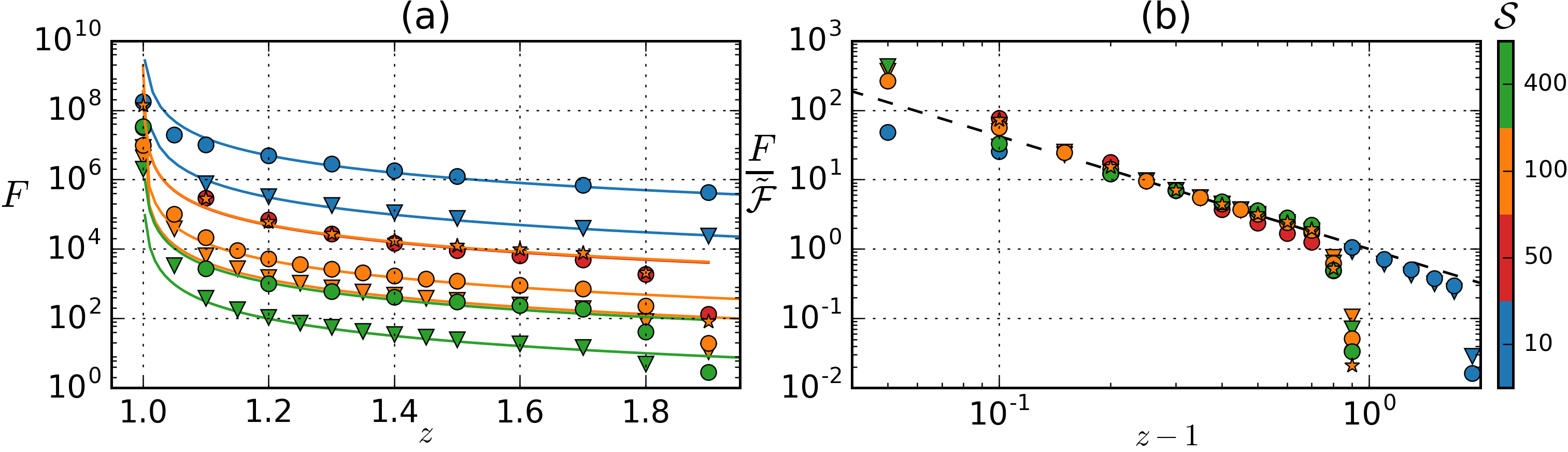}
\vspace{-0.05in}\caption{(a) Total flux in the wave region in DNS (triangles, circles, stars show results for $Ra=4\times 10^7,2\times 10^8,10^9$) along with the predictions $F^{th}$ (solid lines) using the best-fit values for $\mathcal{A}$ and $f_c$ reported in table \ref{table1}. (b) The normalized flux $F/\tilde{\mathcal{F}}$ collapses for all simulations on the prediction $F^{th}/\tilde{\mathcal{F}}\approx (z-1)^{-13/8}$ (dashed line) for $(z-1)\in[0.1,0.8]$ (cf. Eqn \eqref{eq:Fasym}).}\label{fig6}
\end{figure}

We obtain $f_c$ for each simulation case by requiring that the asymptotic prediction for the total flux \eqref{eq:Fasym} equals the total flux in DNS at some height $z^*$ in the wave region, i.e. 
\ba{}\label{eqn:best-fit f_c}
f_c^{\f{21}{2}} =  \f{F(z^*)}{\mathcal{A}}f_N^{\f{47}{8}} \f{28(8\pi^2)^{\f{13}{8}}}{5\Gamma\lp 5/8\rp } \lp z^*-z_i\rp^{{\f{13}{8}}}.
\ea 
This definition of $f_c$ may appear degenerate since $f_c$ can take different values depending on $\mathcal{A}$ or $z^*$. However, as discussed below (cf. also figure \ref{fig6}a), once $\mathcal{A}$ is chosen, $f_c$ does not depend on $z^*$ provided that $z^*$ is sufficiently far from the interface (i.e. $z^*>1.1$). With $\mathcal{A}\sim\co(10)$, we find for all simulations that $f_c$ is almost equal to $u_{rms}$ (to within a factor of 2), which can be considered as a reference value for the turnover time of the largest-scale eddies \cite[][]{lecoanet2015}. From now on we will set the multiplicative constant arbitrarily to $\mathcal{A}=30$. Other values could be chosen, but do not lead to significant changes as long as $f_c$ is relatively close to $u_{rms}$.
 
We compare DNS results for $F(z)$ (symbols)  with the theoretical prediction $F^{th}(z)$ (solid lines) in figure \ref{fig6}(a) for the eight simulation cases, using the parameter values listed in table \ref{table1} (i.e. with $\mathcal{A}=30$). The total wave flux decreases with $z$ in all cases and, as expected, is larger for higher $Ra$ (for fixed $\mathcal{S}$) and lower $\mathcal{S}$ (for fixed $Ra$) \cite[][]{couston2017}. With $f_c$ obtained from a best-fit at just one particular height, we find an excellent agreement between DNS and the theoretical prediction in all of the bulk  of the wave region ($z\in[1.1,1.8]$) for all simulations, which demonstrates that $f_c$ would be similar for any $z^*\in[1.1,1.8]$. The $(z-1)^{-13/8}$ scaling predicted by the asymptotic solution is well verified in figure \ref{fig6}(b), which shows the normalized wave flux $F/\tilde{\mathcal{F}}$ (cf. equation \eqref{eq:Fasym}). All DNS results collapse on the asymptotic solution $(z-1)^{-13/8}$ shown by the dashed line for $(z-1)\in[0.1,0.8]$, demonstrating that the theoretical flux spectrum integrated over the full range of waves excited compares well with the total flux in DNS.

\subsection{Wave energy flux spectrum}\label{sec:discussion2}

\begin{figure}
\centering
\includegraphics[width=1\textwidth]{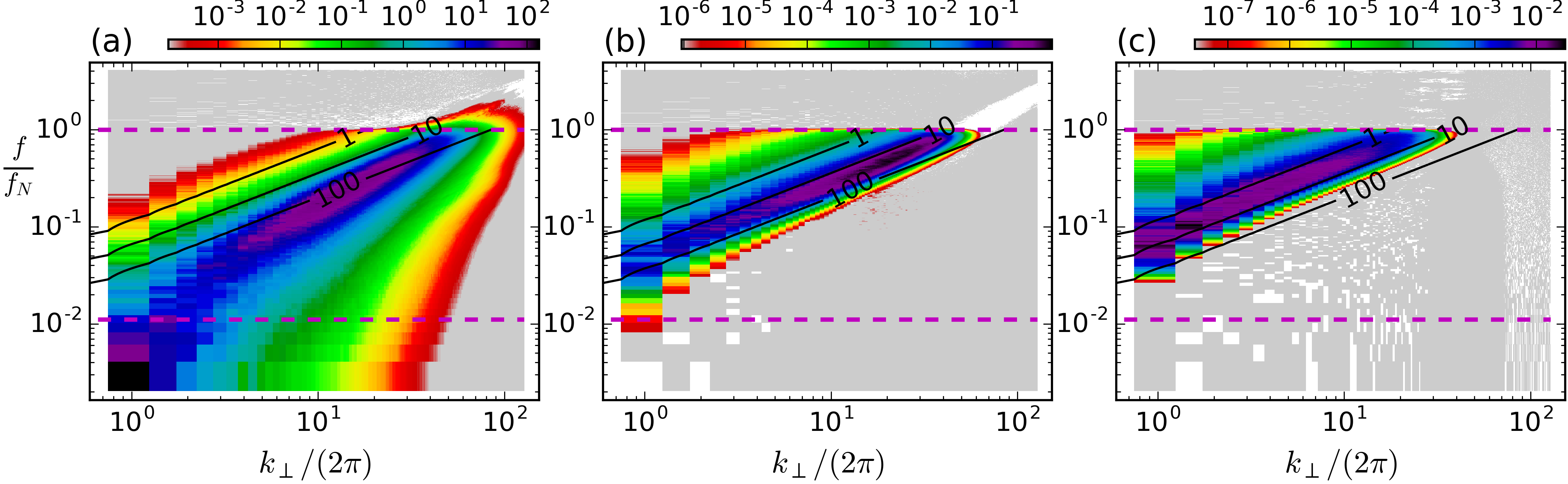}
\vspace{-0.1in}\caption{Wave flux spectrum $k_{\perp}fdF$ in DNS for case $C_9^{100}$ at (a) $z=1.0$, (b) 1.1, (c) 1.5. The color scale is shifted to lower amplitudes from (a) to (c) because the wave flux decreases with height. Solid black lines show isocontours of the decay rate $2\gamma$. Dashed purple lines show $f_c,f_N$. White bins indicate $k_{\perp}fdF$ values that are negative.}\label{fig3}
\end{figure}

We now investigate in details the scalings of the wave flux spectrum $k_{\perp}fdF$ in DNS with $k_{\perp}$ and $f$, and discuss how they compare with the theoretical prediction $\sim k^4f^{-13/2}$. We first show in figure \ref{fig3}(a)-(c) the overall wave flux spectrum $k_{\perp}fdF$ in $(k,f)$ space at heights $z=1,1.1,1.5$ for simulation $C_9^{100}$, which is the most turbulent case. The wave flux at $z=1$ can be difficult to interpret because it contains traces from both waves and overshooting plumes, with one or the other type dominating depending on whether the convection zone extends slightly above or slightly below $z=1$. At $z=1.1,1.5$, however, we are clearly in the wave region and the spectrum has all of the features predicted by the theory: in the high-frequency low-wavenumber range (top left corner), the wave flux spectrum decreases rapidly with frequency and increases with wavenumber up to a cutoff wavenumber. The line of maximum wave flux spectrum overlaps with an isocontour of the decay rate $2\gamma$ (cf. solid black lines and equation \eqref{i}), and that isocontour is closer to the top left corner as $z$ increases because this corner corresponds to waves which are the least damped. The wave flux spectrum falls off quickly at frequencies $f>f_N$ (top dashed line), which is the maximum frequency for propagating waves. Note that while upward-going waves require $dF>0$, few $(f,k_{\perp})$ modes can be seen to have azimuthally-averaged $dF \leq 0$ in figures \ref{fig3} (shown by white bins). However, they have negligible $|dF|$ compared to positive $dF$'s, which is why we restrict the colormap to $dF>0$ only.

\begin{figure}
\centering
\includegraphics[width=1\textwidth]{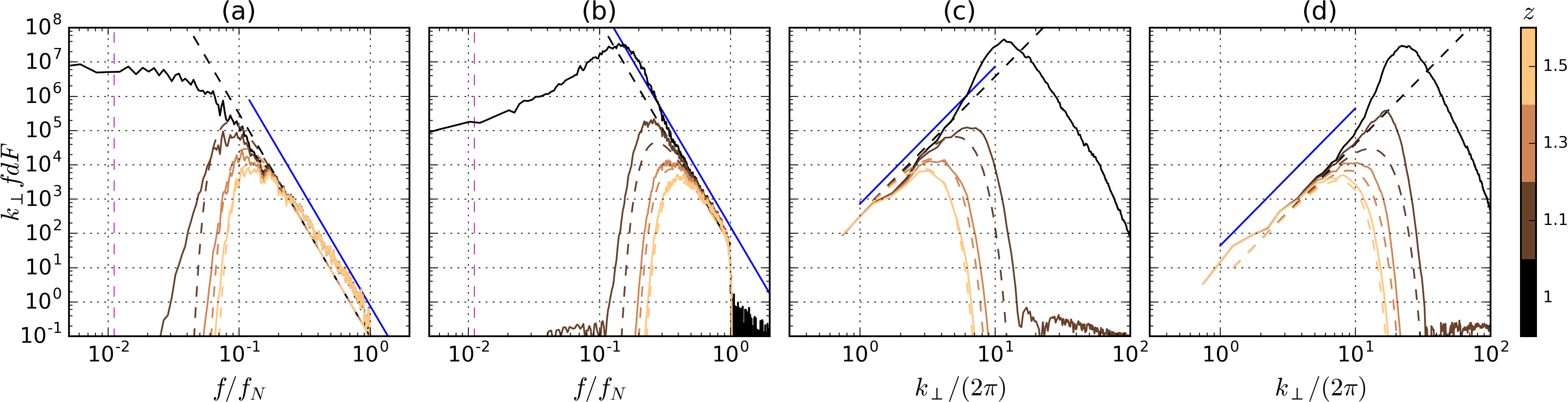}
\vspace{-0.2in}\caption{Wave flux spectrum $k_{\perp}fdF$ for fixed $k_{\perp}/(2\pi)=1.7, 8.2$ [(a)-(b)] and fixed $f/f_N=0.2,0.4$ [(c)-(d)] for case $C_9^{100}$. Lighter colors show results at higher heights. Solid lines show DNS results and dashed lines show theoretical results using the convection parameters listed in table \ref{table1}. The blue lines show the theoretical scalings. }\label{fig4}
\end{figure}

The wave flux spectrum is shown as a function of $f$ for fixed wavenumber in figures \ref{fig4}(a),(b), and as a function of $k_{\perp}$ for fixed frequency in figures \ref{fig4}(c),(d), providing a quantitative comparison of DNS results (solid lines) with the theory (dashed lines). The decrease of the wave flux spectrum in DNS with $f$ is steep and close to the predicted $f^{-13/2}$ scaling (shown by the blue solid line). The increase of the  wave flux spectrum with $k_{\perp}$ closely follows the scaling $k_{\perp}^{4}$, also predicted by the theory (shown by the blue solid line). Again, in figures \ref{fig4}(a),(b) [resp. (c),(d)] we note that the maximum flux moves toward higher frequencies (resp. lower wavenumbers) at higher $z$ as a result of damping. We have checked that similar trends in frequency space and wavenumber space are obtained for other low wavenumbers (within the inertial subrange) and high frequencies ($f\gg u_{rms}$), suggesting an overall good agreement between DNS and theory for $C_9^{100}$.

\begin{figure}
\centering
\includegraphics[width=\textwidth]{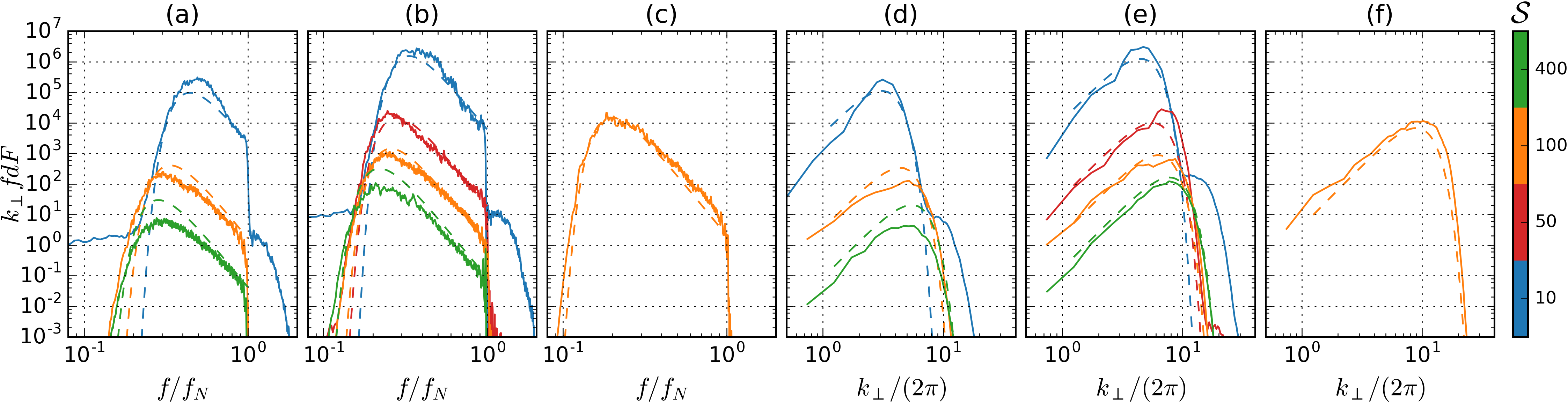}
\vspace{-0.2in}\caption{Wave flux spectrum $k_{\perp}fdF$ at $z=1.3$ for fixed $k_{\perp}/(2\pi)=3.7$ [(a)-(c)] and fixed $f/f_N=0.4$ [(d)-(f)] with $Ra=4\times 10^7,2\times 10^8,10^9$ increasing from (a) to (c) and (d) to (f). The different colors correspond to different stiffnesses $\mathcal{S}$. Solid lines show DNS results and dashed lines show theoretical results using the convection parameters listed in table \ref{table1}.}\label{fig5}
\end{figure}

We now demonstrate that a good agreement is obtained between DNS and the theory not only for $C^{100}_9$ but for all simulations listed in table \ref{table1}. In figures \ref{fig5}(a)-(c), we show the wave flux spectrum at $z=1.3$ as a function of $f$ for $k_{\perp}=3.7$ with $Ra=4\times 10^7,2\times 10^8,10^9$ in (a), (b), (c). Clearly, the DNS results (solid lines) overlap well with the theoretical predictions (dashed lines) obtained after estimating the convective frequency $f_c$ from a best-fit approach (cf. \S\ref{sec:discussion3} and listed values in table \ref{table1}). A good agreement is similarly obtained for the wave flux spectrum at $z=1.3$ as a function of $k_{\perp}$ and for $f/f_N=0.4$ between DNS and the theory (cf. figures \ref{fig5}(d)-(f)). We have checked that similar overlaps between DNS and theory are obtained for other fixed wavenumbers in the inertial subrange and frequencies $f_N> f\gg f_c$.

\vspace{-0.1in}
\section{Concluding remarks}

We have shown that the scaling of the energy flux spectrum of internal waves generated by turbulent convection in DNS follows the  $k_{\perp}^4f^{-13/2}$ prediction in the high $\mathcal{S}\gg 1$ regime \cite[][]{lecoanet2013}, and consistently, the total flux decreases as $(z-1)^{-13/8}$ for all simulations presented. As discussed in the previous section, we can match the flux predicted by the theory to the DNS if we multiply the analytical prediction by a universal constant $\mathcal{A}$ and use a best-fit for $f_c$. It is unsurprising that the theory needs be rescaled since numerical factors have been dropped off along the derivation, and it is equally unsurprising that the best-fit $f_c$ equals a few times the reference frequency based on the rms velocity $u_{rms}$. It is worth pointing out, however, that there are some variations of $f_c/u_{rms}$ across cases, which may mask some effects not included in the theory. Future work should explore the sweeping hypothesis, and include longer integration time with additional simulations or laboratory experiments in the high turbulence, high $\mathcal{S}$ regime.

It is remarkable that there is such a satisfactorily agreement between the DNS and an approximate-at-best theory based on second-order statistics of turbulent motions. The theory, as simple as it is, may prove useful in assessing the detectability of internal waves in geophysical and astrophysical fluids such as stars and in estimating wave-driven changes of global dynamics from the energy in turbulent sources. \\

\textbf{Acknowledgements.} The authors acknowledge funding by the European Research Council under the European Union's Horizon 2020 research and innovation program through Grant No. 681835-FLUDYCO-ERC-2015-CoG. D. L. is supported by a PCTS fellowship and a Lyman Spitzer Jr fellowship. Computations were conducted with support by the HPC resources of GENCI-IDRIS (Grant No. A0020407543 and A0040407543) and by the NASA High End Computing (HEC) Program through the NASA Advanced Supercomputing (NAS) Division at Ames Research Center on Pleiades with allocations GID s1647 and s1439.

\appendix

\section{}\label{appA}
Here we present the key steps required to derive \eqref{eq:Fasym}. Let us use the variables $f/f_c=\xi$, $k/k_c=\eta$, $2(2\pi)^2[k_c/(2\pi)]^3f_N^3f_c^{-4}(z-z_i)=z^*$. The integration of \eqref{eq:Fasym} in $\eta$ results in
\ba{}\notag
& \f{F(z)}{u_c^3\f{f_c}{f_N}} = \int_{1}^{\f{f_N}{f_c}} \int_{1}^{\xi^{\f{3}{2}}}   \f{\xi^{-\f{13}{2}}\eta^4}{\xi\eta} e^{-\eta^3\xi^{-4}z^*} d\eta d\xi =  \int_{1}^{\f{f_N}{f_c}} \xi^{-\f{15}{2}} \int_{1}^{\xi^{\f{3}{2}}}  \eta^3 e^{-\eta^3\xi^{-4}z^*} d\eta d\xi \\ \notag
 & = \int_{1}^{\f{f_N}{f_c}} \biggl\{ \underbrace{   \f{1}{3z^*} \lb \xi^{-\f{7}{2}}e^{-\xi^{-4}z^*}-\xi^{-2}e^{-\xi^{\f{1}{2}}z^*} \rb    }_{\mathcal{I}_1} + \underbrace{ \f{\xi^{-\f{13}{6}} }{9(z^*)^{\f{4}{3}}} \lb \Gamma\lp\f{1}{3},\xi^{-4}z^*\rp-\Gamma\lp\f{1}{3},\xi^{\f{1}{2}}z^*\rp \rb  }_{\mathcal{I}_2} \biggr\}  d\xi,
\ea
with $\Gamma$ the upper incomplete gamma function. Integrating $\mathcal{I}_1$ with respect to $\xi$ yields
\ba{}\notag
\mathcal{I}_1 &= \biggl[ \f{\Gamma\lp\f{5}{8},z^*x^{-4}\rp}{12(z^*)^{\f{13}{8}}}  \biggr]_1^{\f{f_N}{f_c}} - \biggl[ \f{z^*}{3}\text{Ei}\lp -z^*x^{\f{1}{2}}\rp +\f{e^{-z^*x^{\f{1}{2}}}\lp z^*x^{\f{1}{2}}-1 \rp}{3z^*x} \biggr]_1^{\f{f_N}{f_c}}  \approx \f{\Gamma\lp\f{5}{8}\rp}{12(z^*)^{\f{13}{8}}}
\ea
with $\text{Ei}$ the exponential integral and the final expression is the leading-order approximation in the limits $f_N/f_c\gg 1$, $z^*\gg 1$ and $z^*(f_c/f_N)^4\ll 1$. In the same limits, $\mathcal{I}_2 \approx \f{2}{21}\Gamma(5/8)(z^*)^{-13/8}$, such that adding the leading-order approximations for $\mathcal{I}_1$ and $\mathcal{I}_2$ yields equation \eqref{eq:Fasym}. Equation \eqref{eq:Fasym} is  valid for roughly all simulation parameters and $(z-z_i)\in[0.01,10]$.

\bibliographystyle{jfm}
\bibliography{Main.bbl}

\end{document}